\documentstyle[11pt,newpasp,twoside,epsf]{article}
\def\edcomment#1{\iffalse\marginpar{\raggedright\sl#1\/}\else\relax\fi}
\reversemarginpar

\begin{document}
\title{Microquasars}
\author{Marc Rib\'o}
\affil{Service d'Astrophysique, CEA Saclay, B\^at. 709, L'Orme des Merisiers, 91191 Gif-sur-Yvette, Cedex, France}

\begin{abstract}
I partially review our current knowledge on microquasars, making special
emphasis on radio interferometric observations, and I comment on emerging
trends in this field of astrophysics.
\end{abstract}

\section{X-ray binaries, radio emission and microquasars}

An X-ray binary is a binary system containing a compact object, either a
neutron star or a stellar-mass black hole accreting matter from the companion
star. The accreted matter carries angular momentum on its way to the compact
object, around which, in most cases, forms an accretion disk responsible for
the X-ray emission. The most recent catalogs contain a total of 280 X-ray
binaries (Liu, van Paradijs, \& van den Heuvel 2000; Liu, van Paradijs, \& van
den Heuvel 2001). In 131 of such systems the companion star has a spectral
type O or B, and they are classified as High Mass X-ray Binaries (HMXBs),
where mass transfer takes place via a decretion disk (in Be stars) or via a
strong stellar wind or Roche-lobe overflow (in OB supergiants). In 149 X-ray
binaries the optical companion has a spectral type later than B, and they are
called Low Mass X-ray Binaries (LMXBs), where mass transfer occurs via
Roche-lobe overflow.

Among them, this author has found 43 Radio Emitting X-ray Binaries (REXBs), of
which 8 are HMXBs that usually show persistent radio emission and 35 are
transient LMXBs. Since the detected radio emission (in nearly all cases)
displays nonthermal spectra, shows some degree of polarization, and implies
high brightness temperatures, it is interpreted as produced by the synchrotron
radiation mechanism, which takes place when we have charged particles
accelerated in the presence of magnetic fields.

A microquasar is simply a REXB displaying relativistic radio jets that can be
imaged at a variety of angular scales using different interferometers. Due to
the lack of space, I refer the reader to Mirabel \& Rodr\'{\i}guez (1999) and
to Fender (2004) for extended reviews on the topic (including discussion on
emission mechanisms, relativistic effects, observed radio/X-ray correlations,
energetics, jet formation, etc.). The name microquasar was given not only
because of the observed morphological similarities between these sources and
the distant quasars but also because of physical similarities, since when the
compact object is a black hole, some parameters appear to scale with the mass
of the central object. In this context, the temperatures of the inner parts of
the accretion disks are of the order of $\sim$$10^7$~K in the case of
microquasars containing stellar-mass black holes and $\sim$$10^5$~K in the
case of quasars containing supermassive black holes
($10^6$--$10^9$~$M_{\sun}$). This explains why in microquasars the accretion
luminosity is radiated in X-rays, while it is done in the optical/UV domain in
the case of quasars, and why we had to wait until the era of high-energy
astrophysics to discover these sources. On the other hand, the characteristic
jet sizes seem to be proportional to the mass of the black hole, since radio
jets in microquasars have typical sizes of the order of light years, while
radio jets in quasars reach distances up to several million light years in
giant radio galaxies. Last, but not least, the timescales are also directly
scaled with the mass of the black hole following $\tau\simeq R_{\rm
Sch}/c=2GM_{\rm X}/c^3\propto M$, being $R_{\rm Sch}$ the Schwarzschild
radius. Therefore, phenomena that take place in timescales of years in quasars
can be studied in minutes in microquasars. In this sense, one can say that
microquasars mimic, on smaller scales, many of the phenomena seen in AGNs and
quasars, but allow a better and faster progress to understand the
accretion/ejection processes that take place near compact objects. However, it
should be noted that the angular scales in units of $R_{\rm Sch}$ are much
higher for nearby quasars than for microquasars, allowing a {\it closer} look
to the jet formation region.

The number of currently known microquasars is around 16 (Rib\'o 2002), among
the 43 catalogued REXBs. It is interesting to note that some authors (e.g.,
Fender 2004) have proposed that all REXBs are microquasars (i.e., radio
emission is always in the form of relativistic jets), and would be detected as
such provided that there is enough sensitivity and/or resolution in the radio
observations (some of them conducted in the past). In fact, we have detected
radio jets in all REXBs for which detailed observations have been possible,
with the only exception of CI~Cam. Moreover, it has been suggested that all
X-ray binaries, except X-ray pulsars that disrupt the inner part of the
accretion disk where the jets are supposed to be launched, have relativistic
jets (Fender 2004). In this context, all of them should be considered
microquasars. Although this is a reasonable hypothesis, only deep radio
observations of the faint X-ray binary population will allow to confirm or
reject it. The planned improvement of existing radio interferometers will
definitely help to solve this important issue.

\section{Types of jets}

Radio jets have been imaged with different radio interferometers (including
the VLA, VLBA, EVN, MERLIN, ATCA, SHEVE and LBA). Broadly speaking, we can
find the following types of jets.

$\bullet$ {\bf Compact jets.} These jets have been resolved mainly with the
use of the VLBA. Two examples of AU-scale jets are shown in Fig.~1. In the
case of GRS~1915+105 the compact jet appears in the so called `plateau' state
(see Dhawan, Mirabel, \& Rodr\'{\i}guez 2000 and Fuchs et~al.\ 2003). In
Cygnus~X-1 the compact jet is present during the canonical low/hard black hole
state (Stirling et~al.\ 2001). This type of jet displays a flat or inverted
radio spectral index ($\alpha\geq0$, where $S_{\nu}\propto\nu^{+\alpha}$), as
one would expect from optically thick synchrotron emission in a continuous jet
flow (see Fender 2001 and references therein). Therefore, its presence has
been inferred in other black hole candidates where high resolution imaging is
not possible due to their flux density and position in the sky, like in the
case of GX~339$-$4. The flat spectrum has been shown to extend up to mm, IR
and probably optical wavelengths, and even the X-ray emission could be
interpreted as optically thin synchrotron emission produced at the base of the
jet. This is suggested by the X-ray/radio correlation spanning more than three
orders of magnitude in luminosity found in the low/hard state (see Gallo,
Fender, \& Pooley 2003 and references therein).

\begin{figure}
\plotfiddle{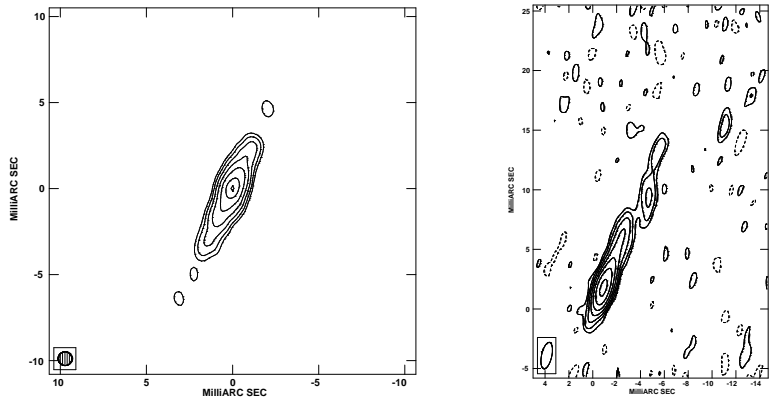}{5cm}{0}{55}{55}{-170}{-160}
\caption{VLBA images of compact radio jets. Left: GRS~1915+105 (from Dhawan et~al.\ 2000). Right: Cygnus~X-1 (from Stirling et~al.\ 2001).}
\end{figure}

$\bullet$ {\bf Discrete ejections.} These ejections take place during `state
transitions' in black hole candidates. The most representative examples are
the superluminal ejections of GRS~1915+105 (Mirabel \& Rodr\'{\i}guez 1994;
Fender et~al.\ 1999) and GRO~J1655$-$40 (Hjellming \& Rupen 1995; Tingay
et~al.\ 1995). In these cases the radio emission rapidly evolves to an
optically thin spectrum (negative spectral index $\alpha$), in agreement with
adiabatic expansion of the relativistic electron clouds. Although the Lorentz
factors are not well constrained due to uncertainties in the distance to the
sources, it has been shown that they have $\Gamma\geq2$. It should be noted
that the neutron star Sco~X-1 shows isolated ejections `moving' at
$\beta\simeq0.45$ ($\Gamma\simeq1.1$) that experience a rebrightening
compatible with an underlying jet flow with $\beta\geq0.95$ ($\Gamma\geq3$)
(Fomalont, Geldzahler, \& Bradshaw 2001). The moving components are
interpreted in this case as the result of jet-ISM interaction. On the other
hand, Cygnus~X-3 appears to display a one-sided relativistic jet at VLBI
scales (Mioduszewski et~al.\ 2001) that may slow down due to interaction with
the ISM and end up as a discrete (isolated) two-sided mildly relativistic jet
(Mart\'{\i}, Paredes, \& Peracaula 2000; Mart\'{\i}, Paredes, \& Peracaula
2001).

$\bullet$ {\bf Large-scale jets.} Parsec-scale radio jets have been imaged in
the two galactic center hard X-ray sources, namely 1E~1740.7$-$2942 and
GRS~1758$-$258 (Mirabel et~al.\ 1992; Mart\'{\i} et~al. 2002). Since these
sources have been found most of times in the low/hard state, the large-scale
jets are thought to result from the long-term action of steady jets on the
ISM. On the other hand, the radio nebula W50 around the well-known microquasar
SS~433 is clearly distorted by the interaction of the jet with it, which
produces as well extended X-ray jets that were detected with the ROSAT
satellite. However, one of the most impressive detections of the jet-ISM
interaction are the simultaneous X-ray (Chandra) and radio (ATCA) observations
of decelerating relativistic jets in the microquasar XTE~J1550$-$564 (Corbel
et~al.\ 2002). The detected X-ray and radio emissions, imaged a few years
after a major ejection and located at around 0.6--0.7~pc from the binary
system, are probably produced by synchrotron radiation mechanism, which takes
place when particles are accelerated in an external shock wave originated when
the jet material interacts with the ISM. It should be noted that a similar
event could have been observed, but only at radio wavelengths, in the source
XTE~J1748$-$248 (Hjellming, unpublished). Finally, one of the latest results
on jet-ISM interaction has been the detection with XMM of extended X-ray jets
spanning $\sim$3.5 arcminutes in the X-ray binary 4U~1755$-$338 (Angelini \&
White 2003). Assuming a distance to the source of 4~kpc these X-ray jets
extend up to 4~pc, as do the radio jets of the galactic center source
GRS~1758$-$258.

An interesting case that does not fit in the ones explained above is that of
LS~5039, where a compact radio jet, but showing discrete components, has been
detected at AU-scales (Paredes et~al.\ 2000). Moreover, the radio spectral
index is $\alpha$$\simeq$$-$0.5, indicative of optically thin emission, in
contrast with the flat spectral index detected in the low/hard state of black
hole candidates (the nature of the compact object in this source is not yet
known, although the radial velocity curve supports a neutron star). This
suggests that the compact jet is build up in the superposition of subsequent
discrete ejections, although this is an idea that has still to be checked.

It should also be noted that simultaneous X-ray, IR and radio observations of
GRS~1915+105 reveal that during episodes of rapid disappearance and follow up
replenishment of the inner accretion disk ejection of relativistic plasma
clouds are produced. These ejections can be understood as small-scale analogs
of the more massive discrete superluminal ejecta of the source (Mirabel
et~al.\ 1998). It is interesting to note that similar phenomena have been
recently observed in the quasar 3C~120, but in timescales of years (Marscher
et~al.\ 2002).

Apart from the jet features discussed above, extended equatorial radio
emission has been detected with the VLBA and MERLIN in the microquasar SS~433
(Paragi et~al.\ 1999; Blundell et~al.\ 2001). Although the observed flat
spectral index suggests thermal emission or self-absorbed synchrotron
emission, its nature is not yet well understood.

\section{Astrometry and stellar evolution}

An emerging trend in microquasars is the study of the space velocity of the
systems, that can be related to the SN explosion of the compact object
progenitor. The basic idea is to combine the radial velocity of the system
with accurate proper motions and the distance to the source to obtain the
total space velocity of the system. Once this is known, we can assume a mass
model for the Galaxy, compute the galactocentric orbit of the system and look
for the parent association of the binary system or the related SN remnant. In
both cases this allows to constrain the age of the binary system after the SN
explosion. The observed velocity (relative to the association or remnant if
available) can be compared with the expected one after assuming different
mass-losses in the SN explosion through basic farmulae (in the case of
symmetric SN explosions) or through Monte-Carlo simulations (in the case of
asymmetric SN explosions with kicks). All this information, specially the
mass-loss during the SN explosion, can shed light on the final stages of
stellar evolution.

It is clear that the compact jets in microquasars, imaged with VLBI
techniques, allow a better measurement of the positions (and proper motions)
of the binary systems than do the optical/IR telescopes (at least nowadays).
Therefore, although this approach was used in the past basically with radial
velocity measurements and upper limits to the proper motions, microquasars
allow a much better and faster study. To this author, the most interesting
results obtained up to now are: an age of $\sim$7~Gyr for the system
XTE~J1118+480, that was probably formed in the galactic halo (Mirabel et~al.\
2001); a runaway velocity of 150 km~s$^{-1}$ in the case of LS~5039 (Rib\'o
et~al.\ 2002a), implying an amazingly huge linear momentum of
$\sim$6000~$M_{\sun}$~km~s$^{-1}$ and a huge mass-loss of at least
6~$M_{\sun}$ during the SN explosion; a delayed black hole formation in the
case of GRO~J1655$-$40 (Mirabel et~al.\ 2002) and a prompt black hole formed
in the case of Cygnus~X-1 (Mirabel \& Rodrigues 2003). Similar studies are in
progress for the microquasars LS~I~+61~303 and GRS~1915+105.

\section{Gamma-rays, matter content, neutrinos and ULXs}

Paredes et~al.\ (2000) have suggested that the microquasar LS~5039 could be 
related to the high-energy $\gamma$-ray source 3EG~J1824$-$1514. The proposed
physical interpretation of this emission is that UV photons from the luminous
optical companion experience inverse Compton scattering by the same
relativistic electrons that later, after having lost part of their original
energy due to IC losses, will account for the radio emission in the jets
(Paredes et~al.\ 2002). A similar scenario could be at work in LS~I~+61~303
(Massi et~al.\ 2001, 2004). These findings open up the possibility that other
unidentified EGRET sources could be microquasars. Using these ideas as the
starting point, Kaufman Bernado, Romero, \& Mirabel (2002) have studied the
general problem of $\gamma$-ray emission arising by inverse Compton scattering
of external photon fields by the electrons of the jets in microquasars. In
this context, they have suggested that some of the unidentified variable EGRET
sources could be precessing microblazars.

The matter content of the jets is only known in the microquasar SS~433, where
iron lines from the jet have been spatially resolved with Chandra (see
Migliari, Fender, \& M\'endez 2002 and references therein). Moreover, if the
jets are hadronic we would expect the formation of TeV neutrinos, that could
be detected in the future with detectors of km$^2$-scale effective area
(Distefano et~al.\ 2002). However, some models to explain jet formation only
work for e$^-$-e$^+$ plasma. Therefore, the observational study of jets to
unveil their matter content is of prime importance to better understand the
jet formation mechanism.

UltraLuminous X-ray sources (ULXs) are off-nuclear X-ray sources in external
galaxies, with computed isotropic X-ray luminosities above the Eddington
luminosity for stellar-mass black holes (see e.g. Makishima et~al.\ 2000 and
references therein). ULXs have been interpreted as evidence for
intermediate-mass black holes. However, Kaaret et~al.\ (2003) have recently
shown that the multiwavelength behavior of the ULX 2E~1400.2$-$4108 in
NGC~5408 is consistent with beamed emission of a relativistic jet from a
stellar-mass black hole, supporting the idea that microquasars in external
galaxies could produce ULXs.

\section{A search for new {\it persistent} microquasars}

As discussed above, microquasars offer a unique opportunity to study
accretion/ejection and related phenomena in human timescales. However, the
current number of this kind of sources is not very high, and it would be
desirable to enlarge the sample. Since the approach to discover new persistent
microquasars was successful in the case of the HMXB LS~5039 (Paredes et~al.\
2000; Paredes et~al.\ 2002), these authors started a long-term project to
discover new similar sources based on a cross-identification between the ROSAT
All Sky Survey Bright Source Catalogue and the NRAO VLA Sky Survey. Although
the first results were promising (Paredes, Rib\'o, \& Mart\'{\i} 2002; Rib\'o
et~al.\ 2002b), recent optical spectroscopic observations (Mart\'{\i} et~al.\
2004) reveal that most of the six studied sources, if not all, are
extragalactic quasars.

Therefore, persistent (probably HMXB) microquasars do not appear to be common
objects in the Galaxy. Nevertheless, it should be pointed out that deeper and
harder X-ray surveys and/or deeper radio surveys, to be conducted in the
future, could change the current situation.

Anyhow, these results stress the importance of studying the already known
REXBs, where radio jets have not been yet clearly resolved but are thought to
exist (Fender 2004), to better understand the involved physics. As an example,
I point out that well-known REXBs like Cygnus~X-1 and LS~I~+61~303 were only
found to display jets after several years of detailed observations (Stirling
et~al.\ 2001; Massi et~al.\ 2001; Massi et~al.\ 2004).

\section{Conclusions}

Microquasars allow to gain insight into jet physics (including formation,
matter content, propagation and interaction with the ISM), and on the origin
and evolution of black holes. They could also be sources of high-energy
$\gamma$-rays and TeV neutrinos, and eventually be ULXs. There are still lots
of open questions that hopefully we will be able to answer with the advent of
new instrumentation.

\acknowledgments
The author thanks Jorge Combi, Qingzhong Liu and Sylvain Chaty for their comments after a careful reading of the manuscript.
He also acknowledges support from a Marie Curie individual fellowship under
contract No.~HPMF-CT-2002-02053.

\end{document}